IAC-20-E2.1.9

# ASTrAEUS: An Aerial-Aquatic Titan Mission Profile

**James E. McKevitt**[ab]*

[a] *Department of Aeronautical and Automotive Engineering, Loughborough University, Epinal Way, Loughborough LE11 3TU,* j.e.mckevitt-17@lboro.ac.uk
[b] *Institute for Astrophysics, University of Vienna, Türkenschanzstraße 17, 1180 Wien, Austria,* jamesm20@univie.ac.at
\* Corresponding Author

**Abstract**
Key questions surrounding the origin and evolution of Titan and the Saturnian system in which it resides remain following the *Cassini-Huygens* mission. In-situ measurements performed at key locations on the body are a highly effective way to address these questions, and the aerial-aquatic platform proposed in this report serves to deliver unprecedented access to Titan's northern surface lakes, allowing an understanding of the hydrocarbon cycle, the potential for habitability in the environment and the chemical processes that occur at the surface. The proposed heavier-than-air flight and plunge-diving aquatic landing spacecraft, *ASTrAEUS*, is supported by the modelling of the conditions which can be expected on Titan's surface lakes using multiphysics fluid-structure interaction (FSI) CFD simulations with a coupled meshfree smoothed-particle hydrodynamics (SPH) and finite element method (FEM) approach in LS-DYNA.

## 1. Introduction

The *Cassini-Huygens* mission gave rise to a much more comprehensive understanding of the Saturnian system [1,2], confirming Titan, Saturn's largest moon, as unique in the Solar System for sustaining a nitrogen-based organically rich atmosphere [3,4], a highly active surface at which complex geological processes occur [5] and a subsurface ocean [6,7].

A multi-phase alcanological* cycle is present, active across the surface of the body, which includes hydrocarbon lakes and seas primarily clustered around Titan's northern pole [8,9], as seen in Fig. 1. This makes Titan the only location in the Solar System other than Earth to have bodies of liquid on the surface and when considered alongside the presence of organically-rich dunes [10], aeolian activity [11], fluvial features [5] and cryovolcanic activity [12], Titan can be seen as analogous to the early Earth, and so is naturally of interest to abiogenetic studies.

The *Cassini-Huygens* mission proved a great success, ending with the spacecraft's intentional destruction in Saturn's atmosphere in a Grand Finale on the 15th of September, 2017. Since then, a great number of proposals for aerobots, balloons, atmospheric probes, lake probes and orbiters have been presented. These are explored further in Section 2.

This report proposes an aerial-aquatic spacecraft for use in the further exploration of Titan and as similar vehicles have been proven viable on Earth [13–15] but never been studied in this context, attempts to perform the required analysis to assess the concept's feasibility.

This report, therefore, aims to present a clear case for the further exploration of Titan by such an aerial-aquatic spacecraft, explain the numerical modelling techniques used and justify their selection, present the results of this modelling and then discuss the implications.

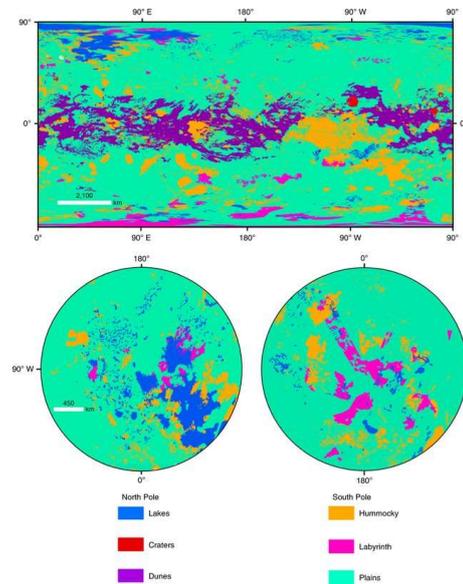

Fig. 1. Mercator map of Titan's major geomorphological units. Reproduced from Lopes *et al.* [9].

---
\* This can be seen as analogous to Earth's hydrological cycle, but is instead driven by methane

### 2. Mission Analysis

Previous mission proposals for Titan show a keen interest of the community to both fund and develop in great detail innovative engineering approaches to further explore the body. Each proposal offers excellent access to one of two mediums at Titan, either surface liquid or solid surface. *Titan Mare Explorer (TiME)* [16] offers the opportunity for limited atmospheric analysis during descent in a similar fashion to *Huygens*, and only offers access to the surface of one of Titan's lakes. NASA's selected *Dragonfly* mission offers the same scientific opportunities as a traditional rover, with a new relocation mechanism. However, following its mission, there will be adequate scope for follow up missions to explore the still unvisited lakes, and the *Titan Saturn System Mission (TSSM)* [17] provides the perfect accommodation for a small and innovative lake lander.

*2.1 ASTrAEUS*

With a unifying aspect of current proposals being their specificity, there is space for a concept which could allow access to multiple mediums with a single platform.

The *ASTrAEUS (Aerial Surveyor for Titan with Aquatic Operation for Extended Usability)* spacecraft (Fig. 2) provides an aerial-aquatic platform inspired by the flight and 'plunge-diving' landing of the gannet sea bird within a field termed bioinspiration and biomimetics.

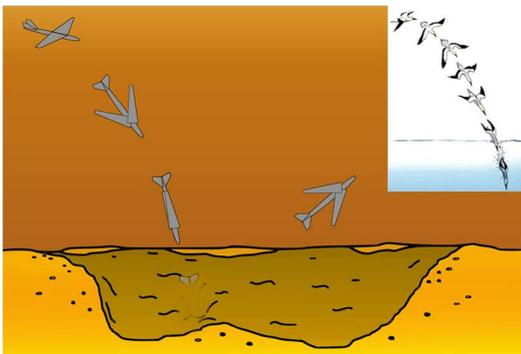

Fig. 2. Impression of a 'plunge-diving' manoeuvre by an aerial-aquatic vehicle inspired by the gannet seabird (inset). Inset adapted from [14].

*2.1.1 Bioinspiration and biomimetics*

The area of bioinspiration and biomimetics is one which studies the natural world using evolution as a guide for the design of new systems to complete processes, in this case, the traversal of two mediums with a single vehicle.

The field is one which has gathered a much-increased following in recent years due to advancements in microrobotics, enabling the replication of the processes which have been observed in nature for many years. Although aerial-aquatics (the study of biological organisms which traverse atmospheric and liquid mediums) has been given more attention recently, to the knowledge of the author it has never before been considered within the context of space exploration.

*2.1.2 Aerial-aquatic operation at Titan*

A key and unique benefit of the *ASTrAEUS* proposal is the ability to make in-situ measurements in various separate bodies of surface liquid on Titan, allowing characterisation of these areas to a level currently unavailable with any planned or proposed mission. Kraken Mare is selected as an initial landing site due to it being Titan's largest lake, and due to its proximity to a number of other lakes (see Fig. 3) of key interest to the scientific community.

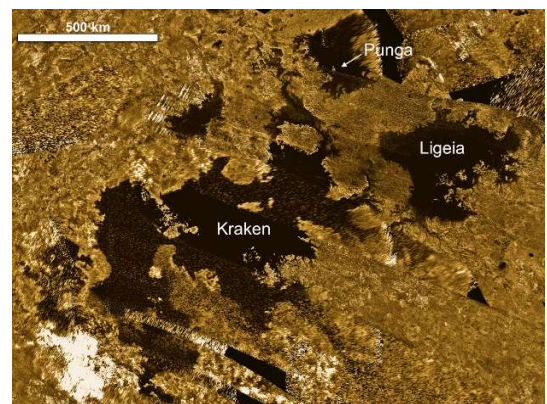

Fig. 3. Cassini SAR mosaic images of the north polar region showing Kraken, Ligeia and Punga Maria. Reproduced from Mitri *et al.* [18].

Just as the *Huygens* probe provided a view to the surface, which was obscured to *Cassini* orbiter, the *ASTrAEUS* aerobot will also give wide-ranging access to the surface and near-surface atmosphere of Titan.

*2.1.3 Science Case*

The opportunities for in-situ study of surface lakes and the near-surface atmosphere above them, given by this vehicle, are unprecedented. Of particular interest in these locations and deducible by measurements made here will not only be the composition of Titan's lakes and near-surface atmosphere, but the methane flux from surface liquid and terrain. It can be hoped that this will aid the development of current understanding, or lack thereof, about the presence of methane at Titan and any sources of resupply – a topic relevant to the discussion about the potential presence of life at the Saturnian body. Key science objectives of the vehicle would include:

1. Characterise Titan's lakes and determine their impact on the hydrological cycle
2. Find the source of Titan's methane
3. Characterise Titan's near-surface atmosphere

### 3. Numerical Modelling

This numerical study involves the CFD simulation of a rigid body entering a finite domain free-surface [†] without heat transfer.

A two-stage approach was taken to first validate modelling techniques, for example, non-reflecting boundaries and equation of state selection – more information on which is detailed in Section 4. This involved using a documented test rig [19,20], which has been refined with experimental data [21], for testing equation of state and other material property changes for the verification of expected behaviour.

Further improvement of fluid behaviour with, for example, bulk viscosity controls can then be put in place in an environment where behaviour is easier to refine.

The test rig was chosen as it not only clearly displays sloshing behaviour, useful for verifying the effects of relative viscosity changes, but it also includes an FEM mesh for FSI modelling. This is useful in refining contact behaviour between materials, as is detailed in Section 4. The test rig was first used to simulate liquid water, and then adapted to the Nitrogen-Ethane-Methane mix found in Kraken Mare, this having density $\rho$ = 664 kg/m$^3$ and dynamic viscosity of $\mu$ = 1014 µPas [22].

For the surface liquid impact, a Computer Aided Design (CAD) model of the *ASTrAEUS* spacecraft was greatly simplified to consist solely of a quasi-ellipsoid fuselage, modelled by a rotated symmetrical NACA-0010 aerofoil. An atmosphere can be neglected by beginning simulations immediately before liquid impact, meaning any drag effects on the projectile would be trivial. SPH particles are also deactivated once they are thrown up just beyond the surface as they cease to be of importance, and so atmospheric interactions with these are irrelevant.

### 4. Methodology
*4.1 Wave-structure interaction*
*4.1.1 Smoothed-particle hydrodynamics*

SPH is a pure Lagrangian method using meshfree particles, as opposed to an Eulerian fixed mesh method. SPH was initially used for astrophysical problems, and now serves to efficiently model situations where large boundary deformations are present (like in this case), or when a free-surface flow needs to be defined. This method serves as the most appropriate for free-surface flow and hydrodynamics problems due to the versatility and simple method of numerical analysis and uses interpolation to compute smooth field variables.

It is widely accepted that for problems with large deflections or boundary deformations, an SPH method is more appropriate than the alternative meshed method as it avoids problems such as mesh distortion, mesh tangling and inaccurate modelling due to unrealistic mesh interdependence.

An SPH method was more readily available to the author at the time of writing, and when beginning the project, large mesh deformations were very possible. It was unknown to what depth the projectile would penetrate the liquid, and so it proved a safe starting point, with FEM always available for adoption later.

The properties of each individual particle are determined by integrating the discretized Navier-Stokes equations according to the physical properties of neighbouring particles which fall within a given range, called the 'smoothing length'. A distance-based function, acting within the smoothing length, weights particles with a closer proximity more heavily to accurately estimate a local density. This works in a similar way to the traditional method of estimating density, that being to divide the total mass inside a given sampling volume by the volume itself. Here, however, to address clustered and sparse regions giving erroneous results, a weighting based on proximity to the sampling volume's centroid helps to smooth the local density estimates.

LS-DYNA requires the definition of an initial number of particles to interact with for each individual particle, so that the smoothing length can be defined according to user requirements. This value was adjusted to improve the accuracy of the simulation, at the obvious cost of computational performance. The accuracy was attained by comparison with the empirical data of Gomez-Gesteira *et al.* [21], and a final value of 500 particles was defined.

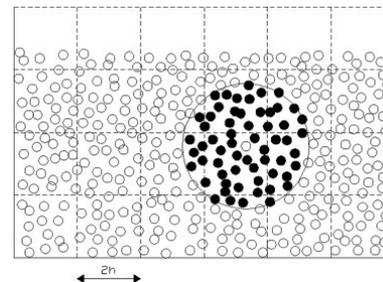

Fig. 4. Set of neighbouring particles in 2D. The possible neighbours of a fluid particle are in the adjacent cells but the particle only interacts with particles marked by black dots. Reproduced from Gomez-Gesteira *et al.* [21].

*4.1.2 Particle approximation theory*

A number of equations are used to define the behaviour of SPH particles. Most notable of these are the equations for energy and momentum, Equ. 1 and Equ. 3 respectively.

---

[†] An undisturbed liquid surface at rest which is subjected to no parallel shear stress

$$\frac{dv^\alpha}{dt}(x_i) = \sum_{j=1}^{N} m_j \left( \frac{\sigma^{\alpha,\beta}(x_i)}{\rho_i^2} A_{ij} - \frac{\sigma^{\alpha,\beta}(x_j)}{\rho_j^2} A_{ji} \right) \quad (1)$$

$$\frac{dE}{dt}(x_i) = \frac{P_i}{\rho_i^2} \sum_{j=1}^{N} m_j \left( v(x_j) - v(x_i) \right) A_{ij} \quad (2)$$

For deployment in LS-DYNA, Equ. (1) can be adapted in this instance to better represent the behaviour of two mediums of significantly different densities. This adaption is seen in Equ. 3.

$$\frac{dv^\alpha}{dt}(x_i) = \sum_{j=1}^{N} m_j \left( \frac{\sigma^{\alpha,\beta}(x_i)}{\rho_i \rho_j} A_{ij} - \frac{\sigma^{\alpha,\beta}(x_j)}{\rho_i \rho_j} A_{ji} \right) \quad (3)$$

This is referred to as the fluid formulation of the momentum equation. Computational simulations were performed using both Equ. 1 and Equ. 3. As it is convention to use the latter for fluid modelling, this was always preferred, and as only a very minor increase in simulation time accompanied this, it was selected.

*4.1.3 Equation of state*
When using an SPH method to simulate liquid behaviour, an equation of state (EOS) is required. This is used to define the hydrostatic or bulk behaviour by calculating pressure as a function of other material properties such as density, energy, and temperature.

While previously it was common to use either a linear polynomial or Gruneisen EOS to model internal liquid behaviour, recent developments in LS-DYNA have allowed a new approach using the Murnaghan EOS [23]. This can be used when assuming little to no compressibility in the subject fluid - sometimes referred to in relevant reading as a quasi-incompressible fluid.

The Murnaghan equation of state defines pressure at a point in the fluid as:

$$p = k_0 \left[ \left( \frac{\rho}{\rho_0} \right)^\gamma - 1 \right] \quad (4)$$

while $k_0$ satisfies:

$$c_0 = \sqrt{\frac{\gamma k_0}{\rho_0}} \geq 10 v_{max} \quad (5)$$

where $v_{max}$ is the maximum expected flow velocity.

$\gamma$ is typically chosen to be around 7 and when the above conditions are satisfied, a low compressibility is maintained while relatively large time steps are allowed.

J. J. Monaghan and A. Kos [24] present an equation to determine the coefficient $k_0$ as

$$k_0 = \frac{100 g H \rho}{\gamma} \quad (6)$$

where $g$ is the acceleration due to gravity and $H$ is a specific height taking the depth of the liquid. Taking the gravitational acceleration at Titan's surface as 1.35 m/s$^2$, the depth of the test tank as 1 m and the liquid density as 664 kg/m$^3$, $k_0$ takes the value of approximately $1.2 \times 10^4$. This compares to a value of approximately $1.5 \times 10^5$ for water in the same conditions.

When performing the verification for stability with Equation 5, it is found that the permitted $v_{max}$ is around 1.2 m/s, as opposed to around 3.2 m/s for a similar water case. Under these conditions, however, the criterion is satisfied for the proven γ value of 7.

When defining the material properties, a density and viscosity is described, which is then related to pressure with the above relationship.

Given the simplicity with which Titan's liquid can be modelled using this EOS, it was used in all simulations.

*4.1.4 Further viscosity control*
The default viscosity parameters assumed by LS-DYNA are typically overly dissipative when a low-viscosity fluid is considered. The dynamic viscosity of the subject liquid is approximately $1 \times 10^{-3}$ Pa s, comfortably qualifying for this designation. There are two main approaches to correct for this, both of which are discussed below.

The bulk viscosity of the fluid can be constrained, applying across the whole domain, and requires the definition of two additional coefficients. These are the quadratic bulk viscosity coefficient and the linear bulk viscosity coefficient, denoted as $Q_1$ and $Q_2$ and taking the recommended values of 0.01 - 0.1 and approximately 0 respectively.

An additional constant $QM$ can be introduced, called the hourglass coefficient. This allows control of the hourglass modes within the fluid which are non-physical, zero-energy modes of deformation resulting in no stress or strain. Hourglass energy is expended internally within the fluid to resist these modes and so is a way the fluid dissipates its energy.

These modes occur when performing one-point integration and as their motion is orthogonal to the plane in which strain calculations are performed, work done by their resistance is not accounted for in the energy equation. When $QM$ is combined with $Q_1$ and $Q_2$, hourglass modes can be controlled.

For the cases considered in this study, bulk viscosity control is sufficient as no significant hourglass nodes are excited in the fluid, given they are primarily present in FEM simulations with high deformation.

*4.1.5 Contact modelling*

Contact modelling is a key element of accurate simulations, and as the boundary deformation of a problem increases, so does the importance of the contact methodology selected.

In LS-DYNA, contacts are modelled in pairs, with nodes belonging to one part acting in a slave role and the nodes belonging to the other in a master role. Each timestep, a search is made for penetrations of the two selected node sets and a force, proportional to the penetration depth, is applied to resist the penetration and move nodes to eliminate the penetration. Called penalty-based contact, restoring forces are calculated as a function of the larger elemental material properties. An alternative methodology called constraint-based contact can also be considered, where the restoring force is calculated as a function of individual nodal mass performing the penetration. These are compared in Equ. 7 and 8 respectively.

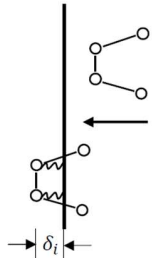

Fig. 5. Slave node-set penetrating master surface, prompting a restoring force

A contact type is now selected. A vast number of options are available in LS-DYNA, the majority of which are not of concern for this problem. For example, surfaces can be specified to react only to specific relative degrees of freedom, allowing sliding interfaces to be defined, or surfaces can be tied together until a specified terminal stress, at which point a fracture forms.

A group of contact types, classified as 'automatic', are a more recently introduced option in LS-DYNA. They are almost always selected as they are capable of handling contacts from any number of directions, making them ideal for handling highly distorted meshes, or as in this case, for handling a large number of individual SPH particles.

For this group of contact types, and many others available, a maximum penetration depth is defined before the slave node's elimination. This acts to ensure restoring forces do not become disproportionate to those experienced in the rest of the model, with the aim of maintaining realistic behaviour.

A different grouping of contact types also separates those which are one-way and those which are two-way. One-way methods identify penetrations of the slave node set within the boundary of the master node set (called the master segment) and apply appropriate restoring forces. Two-way methods, however, identify relevant nodes and apply restoring forces to both node sets, meaning the label of slave and master applied to the node sets in this instance is arbitrary.

As aforementioned, a penalty-based approach to contact modelling has been taken, where a restoring force is a function of penetration depth. This is achieved by the modelling of a spring force, acting between the penetrating node and nearest master segment, with stiffness

$$k = \frac{C_1 \times A^2 \times K}{V} \quad (7)$$

where $C_1$ is the penalty scale factor and $A$ and $K$ are the area and bulk modulus of the contact segment respectively. It should be noted that this equation is only valid for solid elements. An alternative equation is required when shells are considered.

However, an alternative method to calculate $k$ can be to only consider the properties of the individual node. In this case, the contact stiffness is found as a function of the nodal mass where

$$k = C_2 \frac{m}{\Delta t^2} \quad (8)$$

where $m$ is the nodal mass and $\Delta t$ is the timestep. This approach is recommended for impact analysis of dissimilar materials, and after an empirical comparison of both methods, it was confirmed that using Equ. 8 produced the more accurate contact behaviour.

An additional option is available in the software, where further work is done to smooth the contact modelling by interpolating the subject surface. This means considerations like the mesh resolution become less important for low-interest areas, meaning a lower mesh resolution can be used. While it is not recommended to use the method on strongly undulating surfaces, on gentle curves with a uniform structure, this option can be of great help in reducing computational load. While this was considered for application to this problem, access to multiple processors is required if the smoothing is to be used with a two-way contact definition, and this was not available to the author at the time of writing. It can, however, be hoped that this option is available in the future to continue this investigation.

*4.1.6 Material*

For use with the equation of state that has been defined, a further definition of material properties is needed. Fortunately, this proves relatively simple and only the density and dynamic viscosity of the liquid need specifying for use with the Murnaghan EOS. As mentioned above, Titan's liquid has density $\rho$ = 664 kg/m$^3$ and dynamic viscosity of $\mu$ = 1014 µPas, using the assessment in Hartwig *et al.* [22], that Titan's largest lake, Kraken Mare consists of a Nitrogen-Ethane-Methane mix.

*4.1.7 Gravity*

The gravitational acceleration acting on SPH particles and the rigid body is set to 1.35 m/s$^2$ [25].

*4.1.8 Infinite domain modelling*

An intrinsic and highly problematic feature of modelling an infinite domain such as a lake is that in reality, shockwaves caused by a projectile entry will propagate without reflection for a long period, and any reflections that do occur will be weakened to such an extent that by the time of their return, their effect is negligible when compared to other entry effects.

LS-DYNA has recently introduced a new technique to deal with these problems, allowing infinite domains to be modelled in a finite volume, by attempting to eliminate reflections from incident SPH particles. The elimination is achieved by accelerating incident particles parallel to the wall so as to remove them from the path of further incoming particles, without translating the resisting force back into the area of interest. Given the combination of contact models, EOS and material properties, infinite domain modelling attempts resulted in high simulation instability, and non-physical SPH particle velocities. They, therefore, were not implemented into solutions, meaning shockwave reflection time heavily limited results, seen in Section 5.

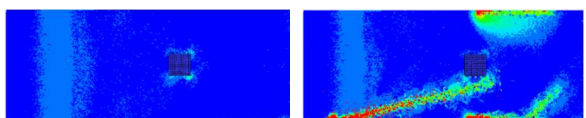

(a) Instant after boundary interaction at time $t = 0$ s  
(b) Boundary instabilities beginning at $t = 0.02$ s

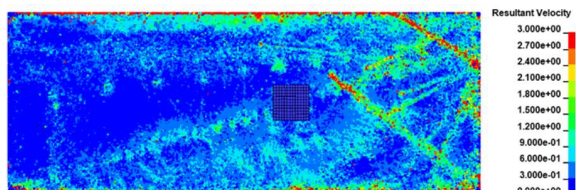

(c) Boundary instabilities fully established at $t = 0.125$ s. $v_{max} > 100$ m/s

Fig. 6. Instability caused by the inclusion of non-reflecting boundaries

*4.2 Projectile impact*
*4.2.1 Resolution and domain size*

The mesh used was generated using LS-PrePost solid meshing using a rotated NACA-0010 aerofoil, and is staged so as to provide a higher resolution at the nose of the structure. The intrinsic problem limited computational power presents to CFD simulations was, as ever, present in this study. However, a condition of the LS-DYNA licence used also meant node-count was limited below 10,000. As it turned out, the requirements of the RAM of the host computer were the limiting factor, and both the resolution of the solid mesh and SPH particle field were responsible for contributing to this.

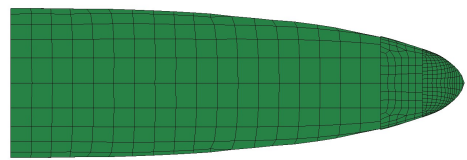

Fig.7. Rotated NACA-0010 trimmed and meshed projectile

Through a mesh sensitivity study, it was seen that a high resolution at the nose was essential. Fig. 8 best represents the problems that are caused by a low resolution at the tip of the projectile. As there are many SPH particles per mesh face, they interact with singular faces in a planar fashion and so produce the non-physical behaviour that can be clearly seen, taking the form of shockwaves propagating not in the form of a uniform wave, but the form of individual particle streams. This effect, however, is only present at the very tip of the nose and is greatly reduced from when an initial global meshing resolution was applied across the whole projectile. Given increased computational power, it would be possible to increase this resolution further and allow for the region of accurate simulation to move forward and closer to the nose. However, that was outside the scope of this study.

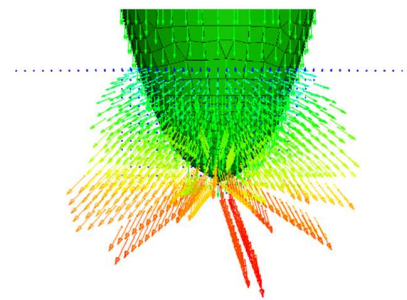

Fig. 8. High SPH resolution, double-precision solved solution with nodes travelling over a specific velocity shown and with attached velocity vectors

Alternatively, a smaller refinement region could be added, in addition to the three already present. However, as can be most clearly seen in Fig. 7, a discontinuity in the mesh is present at boundaries between vastly changing mesh resolutions. This is undesirable at any location, but specifically at a region of high interest. It was concluded through empirical testing that this was unfeasible, and that to eliminate non-physical behaviour caused by mesh resolution, a computer with higher computational power is required.

An alternative method for reducing this effect is to reduce the resolution of the SPH particles, meaning that they react according to multiple FEM faces, as opposed to a single one as is present below around the projectile tip. There are also further reasons for doing this. The simulation below, running on a single PC core (another condition of the software licence available) and with 2133 MHz RAM took 36 hours to progress this small distance into the liquid. The timestep (which is also discussed in more detail below) required for simulation stability for such a high SPH resolution was approximately $1 \times 10^{-7}$ s, and as data was recorded at each timestep to ensure the capture of all simulation detail, produced over 600GB of data. The scale of this simulation, while perhaps expected in industrial applications, is far out of the hardware range of the author, and so the SPH resolution was reduced for the final simulations.

A double-precision solver was also used in this case, with further identical simulations carried out with a single-precision solver. This is recommended practice for initial simulation runs so that results can be compared, and the use of the single-precision solver over the double (the latter having an approximately 30% longer computation time) can be justified. It is also common practice to use the double-precision solver for explosion modelling, and therefore necessary to test it with high-velocity impact studies. Given that the timestep used was small enough, the single and double-precision solvers produced identical results, and so the single-precision solver was used from this point on.

Another problem associated with RAM is not just its speed but its size. The RAM size required to initialise a simulation with SPH particles scales with the domain size and number of particles. Therefore, for an increased domain size, the number of particles must be reduced. A compromise was found between these two competing requirements, whereby the domain size meant the projectile could traverse to a depth sufficient to see a divergence between its deceleration in water and nitrogen-ethane-methane liquid (as seen in results in Section 5), and the SPH resolution meant shockwave propagation was effectively modelled (this being defined by the shockwave travelling at the speed seen in the high-resolution, double-precision simulation).

*4.2.2 Equation of state*
Simply by maintaining the same characteristic length from the wave-structure interaction simulation in the impact simulation, the EOS can remain unchanged. Therefore, the modelled SPH domain is constructed with the same reference dimension. As before, the Murnaghan EOS is used and the parameter $k_0$ is kept the same.

*4.2.3 Timestep resolution*
As previously mentioned, the timestep required to ensure stability in the high resolution, double-precision simulation of Fig. 8 was very small relative to the period of interest for the simulation. This timestep can be specified by adding a scale factor to the calculated time step to ensure a resolution high enough to capture all features. This is however a trial-and-error, empirical process and can be time-consuming. It is also a 'fudge factor' method. A neater approach is to correctly specify contact penalties (as discussed above in Section 4.1.5), which relate to the relative velocities which are seen between our two mediums.

For this simulation, the LS-DYNA computed timestep was around $1 \times 10^{-2}$ s, and as a timestep of $1 \times 10^{-7}$ s was required, a very significant fudge factor would have been required – a particularly bad approach as LS-DYNA's computed timestep changes through the course of the simulation, meaning and small changes are largely amplified, effectively invalidating the simulation.

Following the specification of contact velocities, a relatively small timestep scale factor of 0.6 was required, just below the recommended value for explosive analysis.

## 5. Results and discussion
*5.1 Wave-structure interaction*
*5.1.1 Sloshing*
Fig. 9 shows the simulation of Titan's surface liquid, with identical conditions present for liquid water at identical time steps for reference. While a nitrogen-ethane-methane mix is more viscous than liquid water, this difference is relatively small at $\Delta\mu = 1.24 \times 10^{-4}$. This represents an approximately 15% increase from liquid water. The largest relative change between materials is the density, as the 664 kg/m$^3$ density of N/C$_2$H$_6$/CH$_4$ represents an approximately 35% decrease from that of water.

Both of these key material changes would bring about the behaviour observed in Fig. 9, with viscosity affecting the speed at which the fluid progresses along the container (which can be observed to be much slower), and density affecting the manner of its impact with the structure (as can be seen by the magnitude of the interaction).

An observation, useful for projectile simulations later, about the relative speed and magnitude of energy transfer

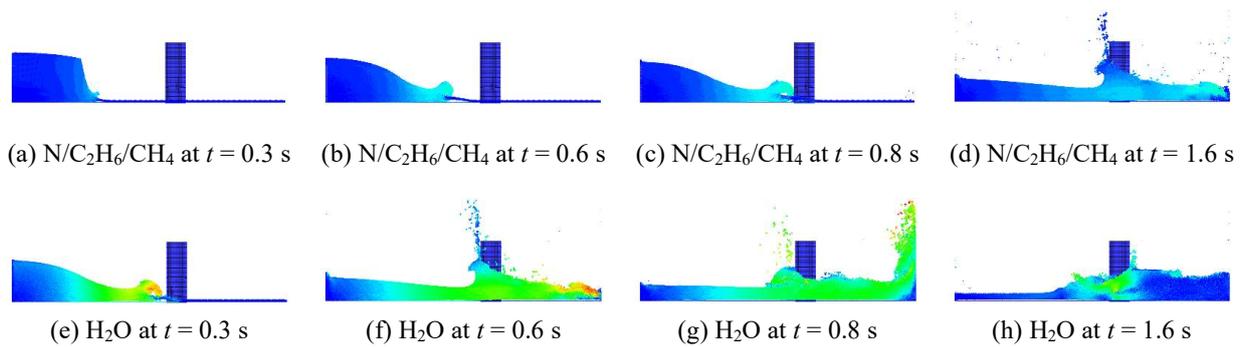

(a) N/C$_2$H$_6$/CH$_4$ at $t$ = 0.3 s  (b) N/C$_2$H$_6$/CH$_4$ at $t$ = 0.6 s  (c) N/C$_2$H$_6$/CH$_4$ at $t$ = 0.8 s  (d) N/C$_2$H$_6$/CH$_4$ at $t$ = 1.6 s

(e) H$_2$O at $t$ = 0.3 s  (f) H$_2$O at $t$ = 0.6 s  (g) H$_2$O at $t$ = 0.8 s  (h) H$_2$O at $t$ = 1.6 s

Fig. 9. Wave development and FSI of Kraken Mare liquid and Earth water with velocity displayed. Red corresponds to a higher relative velocity.

by the fluids can be made. It can be seen that due to the different behaviours, energy is not only transferred to the terminating wall at a slower rate, a smaller magnitude of it is transferred. This is clearly demonstrated in Fig. 12 and is discussed in Section 5.1.3.

*5.1.2 Fluid-structure interaction*

Further interaction with the FEM structure can be seen from above in Fig. 10. In Fig. 10(a), the lower velocity at which the nitrogen-ethane-methane mixture passes the structure means the beginning of a von Kármán vortex street can be seen due to the vortex shedding by the structure. This simulation was not run with the observation of this phenomenon in mind, nor is it the purpose of this study, meaning the container does not provide sufficient space or time for the effect to become fully established. Therefore, while a frequency of oscillation cannot be obtained, it is nonetheless an interesting effect to observe.

It serves as a particularly useful way to observe the changes a different viscosity, density and the subsequent velocity change have on fluid behaviour.

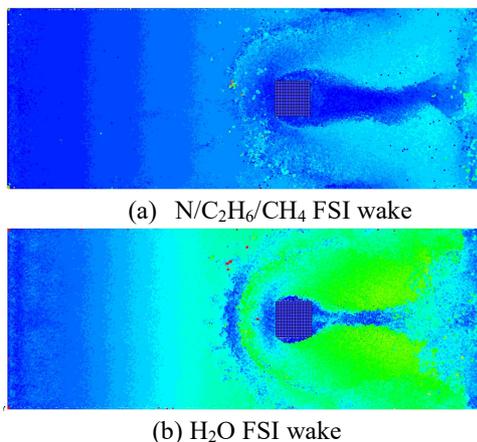

(a) N/C$_2$H$_6$/CH$_4$ FSI wake

(b) H$_2$O FSI wake

Fig. 10. Wave-structure wake formation of Kraken Mare liquid and Earth water with velocity displayed

It is known that turbulence is related to Reynolds number $Re$ in that above a critical value turbulence begins. While this problem is three-dimensional and does not concern a constant flow, this provides interesting context for flow observations.

$$Re = \frac{\rho v d}{\mu} \qquad (9)$$

where $v$ is the fluid velocity, $d$ is the characteristic length and $\mu$ is the kinematic viscosity. With a reduced density, velocity and fractionally increased kinematic viscosity, the Reynolds number is decreased, scaling it to give approximately $Re_{\text{titan}} \approx 0.7 Re_{\text{earth}}$, where $Re_{\text{titan}}$ is the Reynolds number of the N/C$_2$H$_6$/CH$_4$ flow. Therefore, turbulent flow should be even less expected in the Titan case than the Earth case, meaning both simulations take place within the laminar-turbulent $Re_{critical}$ region.

Fig. 11 shows this effect in greater clarity and attaches relevant streamlines.

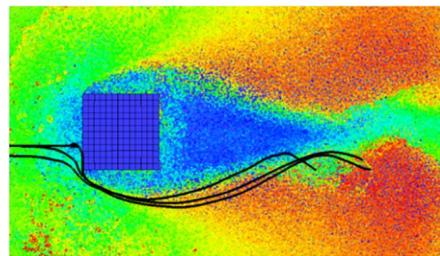

Fig. 11. Vortex shedding for N/C$_2$H$_6$/CH$_4$ liquid with streamlines on particles in area of interest

*5.1.3 Fluid force*

Fig. 12 shows the force on the terminating wall by the respective fluids. As was evident in Fig. 9 the peak force of the water occurs much earlier than the nitrogen-ethane-methane mix. What however was less clear was the relative sharpness of each peak, or in this case the peak of water, as the nitrogen-ethane-methane mix does not appear to demonstrate a peak. It has already been observed that the rate and magnitude of energy transfer across the container was higher for water, but this magnitude can be much more clearly seen in Fig. 12.

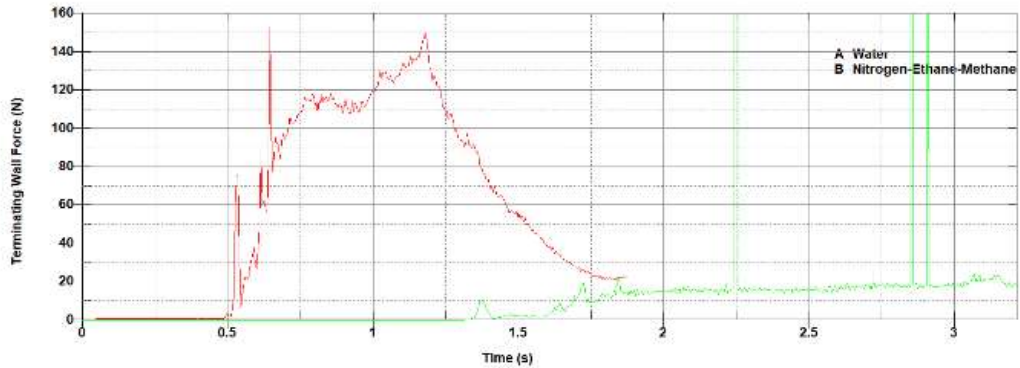

Fig. 12. Force caused by sloshing fluid on the terminating wall for water and N/$C_2H_6$/$CH_4$ mix

*5.2 Projectile impact*
*5.2.1 Shockwave propagation*

One main problem with modelling a quasi-infinite domain in a finite domain is that the solution becomes much less accurate, perhaps to the point of being invalid, once shockwaves caused by the impact are reflected back to the impacting body.

For this simulation, this phenomenon can be visualised by showing the acceleration, within a fine tolerance, of the free-surface liquid. This has been done in Fig. 13, and as expected, the propagating shockwave is generally uniform. The time at which this figure was produced has been chosen to show that the momentum of the incoming projectile pushes the centre of shockwave propagation below the free-surface, as can be seen by the exposed SPH particle wall.

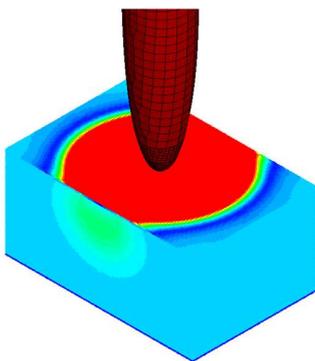

Fig. 13. Shockwave propagation in N/$C_2H_6$/$CH_4$ mix with nodal acceleration displayed

The exact mechanics of shockwave reflection can be more clearly seen in Fig. 14, with the forces on the projectile (causing the plotted deceleration) clearly affected at around $5.75 \times 10^{-3}$ s, the point at which the shockwave returns to the body. Therefore, the figure limits data to that gathered before the shockwave reflection, for the discussion of projectile behaviour. While the figure is cropped to the resultant acceleration values of interest, it should be noted that they exceeded 8 times their peak value of the valid period. Therefore, even if local particles return to enacting accurate physical behaviour after the peak of shockwave reflection, an offset would be needed for the data.

*5.2.2 Projectile behaviour*

The projectile was impacted into the fluid with a velocity of 25 m/s and at an angle of 75° to the horizontal. The object moved under 6 degrees of freedom, and its body was parallel with its initial direction of motion.

The simulation was completed with timesteps of approximately $1 \times 10^{-8}$ s, with data taken at each timestep. The results presented represent a 75-point moving average plot, meaning that data points are plotted at approximately $7.5 \times 10^{-7}$ s intervals. This was done to reduce noise which made the data unintelligible. However, the oscillating forces that the body experienced are still clearly seen. When first observed, these were much larger, and bulk viscosity controls were tightened, in line with the procedure discussed previously. However, it is now understood that constraining hourglass modes would also contribute to damping the oscillations, presenting an area for future work. The oscillations are not discussed further as their presence is erroneous, although it is appreciated that their presence is a cause for further investigation. As discussed above, the reflected shockwave can be seen and so data until this point is what is of interest.

The data shows a similar behaviour of water and the nitrogen-ethane-methane mix initially. The initial increase in deceleration magnitude and its subsequent decline can be attributed to the initial impact on a non-moving body and its attenuation is as expected.

Beginning at approximately $2 \times 10^{-3}$ s, the behaviour of the projectile in the two mediums begins to diverge. The resisting force experienced by the projectile in liquid water begins to become larger than that experienced in the nitrogen-ethane-methane medium.

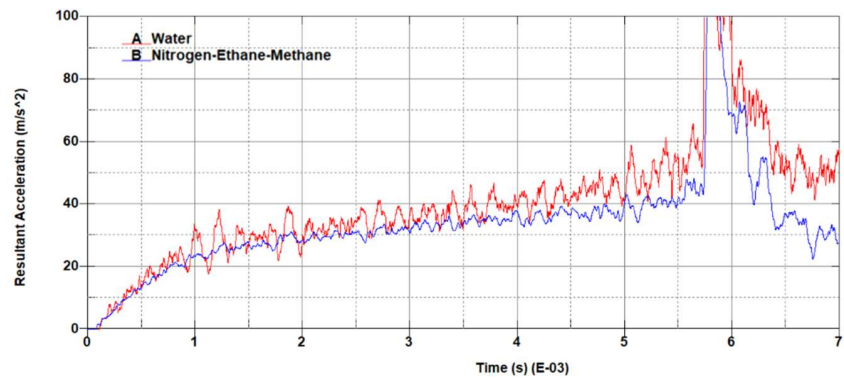

Fig. 14. Projectile deceleration due to water and $N/C_2H_6/CH_4$ mix

Viscosity is a measure of a fluid's resistance to deformation. Therefore, given the two medium's relative viscosities (discussed previously), it could have been expected that nitrogen-ethane-methane would have provided a larger resisting force to the projectile's motion. However, when the densities of the fluids are considered relative to one another, the increased inertia of water is to be expected (it is well known that fluid inertia is directly proportional to density [26]), and it can be clearly seen that this has a larger effect on the projectile's motion than the difference in viscosities.

An additional unexpected observation of the projectile behaviour was made. As seen in Fig. 15, the projectile experienced small, but well established and uniform, oscillations during entry. These occurred with a vector perpendicular to the direction of motion, and in the plane of the projectiles angle to the horizontal with a frequency of 280 Hz.

As mentioned previously, the projectile was orientated in line with its velocity vector, and given six degrees of freedom. However, upon entry, the x and y-components of its velocity were attenuated at different rates, given the geometry of the projectile. Therefore, and oscillating stress within the projectile was excited.

It was not within the initial scope of the project to detect these, and further work needs to be done to properly understand their implications.

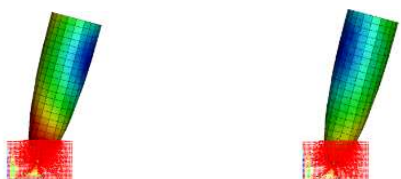

(a) Projectile entry at time $t$
(b) Projectile at time $t + 3.6 \times 10^{-3}$ s

Fig. 15. Oscillation of projectile during fluid entry seen with nodal acceleration displayed

## 6. Conclusions and further work

This work has successfully modelled Titan's surface liquid and presented the results of two key fluid-structure interaction simulations. Using Titan's largest lake Kraken Mare's composition, shown by Hartwig *et al.* [22] to be a nitrogen-ethane-methane mixture and adapting the Murnaghan equation of state and work done by J. J. Monaghan and A. Kos [24], a model has been built.

Using the results of simulations making use of this model, the behaviour of Titan's surface liquid can now be better understood, and the deceleration caused by its liquid can be anticipated. It can also now be understood that despite the higher viscosity, the lower density causes smaller forces on the impact body than those of water.

Taking the simulated projectile impact case, the deceleration of the projectile was very gradual, meaning the projectile could be expected to penetrate the lake to a depth equal to around 25 body lengths given the entry velocity of 25m/s and angle of 75°. This would be ideal given the interest in profiling the lakes of Titan, and appropriate for Kraken Mare, given its estimated depth of 160 m [27].

The mission concept demonstrates key strengths against the other proposals discussed in this paper, given the ability for the *ASTrAEUS* platform to perform almost all of their individual functions, and has certainly inspired interest for future study.

In the future, investigations would benefit greatly from the allocation of more resources such as computational power and licensing capabilities. Given this, more complex geometries could be modelled in a larger domain, increasing simulation accuracy and the length of time for which simulations remain accurate.

Further work should also look at the ejection of the vehicle from the surface liquid. The beginnings of this have already been completed using classical mechanics and hydrodynamics equations, in conjunction with this project, by the author.

In addition to this, higher-level components of space mission proposals should receive some future attention. For example, an analysis of the power requirements of a vehicle such as this would allow the selection of an appropriate power supply such as a Radioisotope Thermal Generator (RTG) and the assessment of whether this would be sufficient, or simply provide enough power for a short release power supply to enable short flight from one body of surface liquid to another. There are a large number of considerations to make at this level, however, and it is recognised by the author that a team would need to be assembled to complete these, another priority for the future.

**Acknowledgements**

This work is the result of a research project funded by the Royal Academy of Engineering, the Royal Astronomical Society and the British Interplanetary Society. Research was conducted in, and with the support of, the Blackett Laboratory at Imperial College London and the Aeronautical Engineering department at Loughborough University. Thanks are given to members of these departments for their belief in this project.